\documentclass[showpacs,amsmath,nofootinbib,amssymb,twocolumn,superscriptaddress]{revtex4-1}
\usepackage{array}
\usepackage{color}
\usepackage{dcolumn}
\usepackage{bm}
\usepackage{epsf}
\usepackage{graphicx}
\usepackage{amsmath} 
\usepackage{dsfont} 

\newcommand{\beq}{\begin{equation}}
\newcommand{\eeq}{\end{equation}}
\newcommand{\bea}{\begin{eqnarray}}
\newcommand{\eea}{\end{eqnarray}}
\newcommand{\nn}{\nonumber \\}

\newcommand\eqn[1]{(\ref{#1})}      
\newcommand\Eqn[1]{Eq.~(\ref{#1})}  

\newcommand{\tr}{\hbox{tr}}

\begin{document}
\topmargin0in
\textheight 8.in 
\bibliographystyle{apsrev}
\title{Neutrino-antineutrino correlations in dense anisotropic media}

\author{Julien Serreau and Cristina Volpe}
\email{serreau@apc.univ-paris7.fr}

\email{volpe@apc.univ-paris7.fr}
\affiliation{Astro-Particule et Cosmologie (APC), CNRS UMR 7164, Universit\'e Denis Diderot,\\ 10, rue Alice Domon et L\'eonie Duquet, 75205 Paris Cedex 13, France}

\begin{abstract}
 We derive the most general evolution equations describing in-medium (anti)neutrino propagation in the mean-field approximation. In particular, we consider various types of neutrino-antineutrino mixing, for both Dirac and Majorana fields,  resulting either from nontrivial pair correlations or from helicity coherence due to the nonvanishing neutrino masses. We show that, unless the medium is spatially homogeneous and isotropic, these correlations are sourced by the usual neutrino and antineutrino densities.  This may be of importance in astrophysical environments such as core-collapse {supernovae}.
\end{abstract}

\date{\today}

\pacs{}

\maketitle

\section{Introduction}

Neutrinos are connected to key open questions regarding fundamental interactions and astrophysical observations. 
Current research questions in neutrino physics include the neutrino masses and mass ordering, the Dirac or Majorana nature of neutrinos, the existence of sterile neutrinos, and leptonic CP violation. 
It is yet to be established how (and how much) neutrinos influence the gravitational explosion of massive stars and  the outcomes
of stellar nucleosynthesis processes such as the $r$--process.  Identifying the answers to this latter question requires the detailed investigation of numerous astrophysical aspects, including an in-depth understanding and accurate treatment of neutrino flavour evolution in dense environments. 

Steady progress has been made since the assessment of the Mikheev-Smirnov-Wolfenstein (MSW) effect \cite{Wolfenstein:1977ue,Mikheev:1986gs} as a solution of the high-energy solar neutrino deficit problem \cite{Robertson:2012ib}. Neutrino self-interaction and dynamical aspects---such as turbulence and the presence of shock-waves---have been shown to induce neutrino flavor changes in the case of massive stars undergoing gravitational core collapse (see, e.g., Ref.~\cite{Volpe:2013kxa}). 

It has long been speculated that neutrino flavor change in matter could impact the supernova dynamics.  While the MSW effect occurs in the outer layers of the supernova, the neutrino-neutrino interaction introduces a nonlinear refraction index \cite{Pantaleone:1992eq} that produces a flavor change in deep regions near the neutrinosphere. Assessing the impact of neutrino self-interactions requires further investigation since their effects might be diluted (or even suppressed) due to decoherence effects in realistic simulations (see, e.g., Refs.~\cite{Duan:2010bg,Duan:2009cd} for reviews and Ref.~\cite{Akhmedov:2014ssa}). Quantitative evaluations of the impact of neutrino flavor changes on shock waves seem to indicate that the conversion occurs outside the gain region and does not affect the explosion \cite{Dasgupta:2011jf}. Moreover, the neutrino  self-interaction might impact the abundance of heavy elements  in neutrino-driven winds in {supernovae} \cite{Balantekin:2004ug,Duan:2010af}, in black hole accretion-disks, or in neutron star--neutron star mergers \cite{Malkus:2014iqa}. These studies and the predictions for supernova neutrino signals are usually made by assuming that the interaction of neutrinos within the star can be accounted for by an effective Hamiltonian and evolution equations that are valid in the mean-field approximation. On the other hand, neutrino transport within the neutrinosphere is based on a Boltzmann treatment for particles without mixings \cite{Janka:2012}.  Exploring the validity of the mean-field approximation is one of the  necessary steps to put  investigations of neutrino flavor conversion and their impact on solid ground.  This requires studying the role of two-body correlations from the high-density region, where neutrinos are trapped, to the low-density regime, where they start free-streaming.  

Equations describing in-medium neutrino propagation have been derived using various approaches either at the mean-field level \cite{Qian:1994wh,Sawyer:2005jk,Balantekin:2006tg,Volpe:2013uxl}, or including collisions \cite{Dolgov:1980cq,Sigl:1992fn,McKellar:1992ja,Vlasenko:2013fja}. 
In the context of supernova physics, various evolution equations beyond the mean-field approximation have been obtained using different theoretical frameworks. Reference\cite{Balantekin:2006tg}  showed that the neutrino mean-field equations---including the neutrino-matter and neutrino-neutrino interactions---correspond to a saddle-point approximation for the coherent state path integral representation of the evolution operator. Implicit corrections beyond the saddle-point approximation were also given.
References \cite{Pehlivan:2011hp,Pehlivan:2014zua} used an algebraic approach to show that the full many-body problem associated with the 
neutrino self-interaction Hamiltonian (without the matter term) is exactly solvable. This Hamiltonian is related to the (reduced) one describing Bardeen-Cooper-Schrieffer (BCS) superconductivity \cite{Pehlivan:2011hp}. In Ref. \cite{Cherry:2012zw}, rescattered neutrinos were considered in a schematic way to mimic a few collisions outside the neutrinosphere, showing that a small fraction of backscattered particles can produce modified flavor patterns. Collisions were included in Ref.~\cite{Vlasenko:2013fja} where the two-particle-irreducible (2PI) effective action formalism \cite{Cornwall:1974vz} was employed to derive evolution equations for
the two-point correlation function.

A different class of corrections to the standard mean-field equations arises from considering various nontrivial two-point correlations beyond the usual neutrino and antineutrino density matrices in flavor space.  
Specific particle-antiparticle correlations for Majorana neutrinos---corresponding to quantum-mechanical coherence between states of opposite helicities---can be present, e.g., due to the coupling of a nonzero neutrino magnetic moment to intense magnetic fields in supernov{ae} \cite{Dvornikov:2011dv,deGouvea:2012hg} or from the neutrino masses.
Numerical results show that the former can modify the neutrino fluxes and produce neutrino-antineutrino conversion, even for a Standard Model magnetic moment \cite{deGouvea:2012hg}. 
These correlations also naturally arise in nonpolarized media when contributions from the nonzero neutrino masses are taken into account, as was observed in Refs.~\cite{Vlasenko:2013fja,Cirigliano:2014aoa}, where generalized kinetic equations implementing helicity coherence were derived.
 
Contributions to the mean-field Hamiltonian proportional to powers of the neutrino masses were first discussed in Ref. \cite{Volpe:2013uxl}, where 
the possible relevance of two-point correlators arising from neutrino-antineutrino pairing correlations has been pointed out.
These have been included in extended mean-field equations using first-principle quantum field-theoretical techniques based on systematic truncations of the Born-Bogoliubov-Green-Kirkwood-Yvon (BBGKY) hierarchy\footnote{Note that such techniques allow one to consistently include collision processes; see, e.g., Refs.~\cite{WC85,Volpe:2013uxl}.} \cite{BBGKY}. Contributions from the neutrino mass to abnormal neutrino-antineutrino correlators were also discussed.
The linearized version of these extended equations was derived in Ref.~\cite{Vaananen:2013qja} using methods from the theory of atomic nuclei and metallic clusters, establishing a link between the collective stable and unstable modes of these many-body systems and those of a gas of self-interacting neutrinos in a dense medium.

Similar two-body pairing correlations play a nontrivial role in various areas of physics, from atomic nuclei to condensed matter and astrophysical systems such as neutron stars. In strongly interacting systems, e.g.,  atomic nuclei, pairing between equal like particles (protons or neutrons) and unlike particles (neutrons and protons) 
are of key importance in correctly determining the properties of the ground state and of the excited states and in describing nuclear dynamics.
In the original BCS theory, the long-range electron-phonon attractive interaction produces the
electron Cooper pairs responsible for superconductivity \cite{Bardeen:1957mv}. In relativistic theories, the possible relevance of particle-antiparticle pairing correlations were pointed out in the context of lepto-/baryogenesis in Refs.~\cite{Herranen:2010mh,Fidler:2011yq}.

In the present work, we derive the most general mean-field equations for Dirac and Majorana neutrinos including all possible types of two-point correlators. The corresponding evolution equations involve nontrivial particle-antiparticle mixing terms. We first give expressions that are valid for a general inhomogeneous system of self-interacting massive neutrinos. We then specify our equations to the case of a spatially homogeneous system and we discuss pairing correlations and helicity coherence terms separately. We treat the former in the limit of ultrarelativistic neutrinos. For the latter, we include nonrelativistic corrections since they arise from  the nonvanishing neutrino masses. We also discuss the conditions for nonvanishing neutrino-antineutrino mixing contributions. In particular, this requires spatial anisotropies of the matter and/or neutrino backgrounds. We derive the mean-field Hamiltonian for a typical astrophysical environment made of electrons, nucleons, and (anti)neutrinos. In general, the latter depend on both the neutrino and antineutrino densities and on the various neutrino-antineutrino correlations. 
We compare our findings with the previous results of Refs.  \cite{Volpe:2013uxl,Vlasenko:2013fja,Cirigliano:2014aoa}.

The paper is structured as follows. Our procedure and general mean-field equations for massive neutrinos in an inhomogeneous background are given for Dirac and Majorana fields in Sec.~\ref{sec:evol}. In Sec.~\ref{sec:pair}, we focus on the role of pair correlations, specializing to spatially homogeneous systems with ultrarelativistic neutrinos; while helicity coherence is investigated in Sec.~\ref{sec:hel}. We conclude in Sec.~\ref{sec:concl}. General equations combining pairing correlations and helicity coherence are presented in Appendix \ref{appsec:total}.

\section{General mean-field equations}
\label{sec:evol}

At any time, the (quantum) state of the system of interest---here, the neutrino gas---can be fully characterized by the values of equal-time correlators of field operators. In general, the latter obey an infinite set of coupled evolution equations, i.e., the quantum field theory generalization of the BBGKY hierarchy \cite{BBGKY,Volpe:2013uxl}. Alternatively, these can be obtained from unequal-time correlators, which also satisfy an infinite set of coupled integrodifferential equations known as the Kadanoff-Baym equations \cite{KB}. In any case, lower-order correlators depend on higher-order ones and one needs to close the hierarchy in one way or another. Powerful systematic approximation schemes can be based on 2PI functional techniques, where $n$-point correlation functions are expressed in terms of the full propagator, which is to be determined self-consistently \cite{Berges:2004vw}. In particular, this provides a convenient starting point to derive kinetic (e.g., Boltzmann) equations when a gradient expansion is justified  \cite{Calzetta:1986cq,Vlasenko:2013fja}.

The mean-field, or Hartree-Fock approximation is the simplest nontrivial closure of such hierarchies. Physically, it corresponds to the propagation of the (anti)particles of interest in the averaged background field generated by the particles of the medium (possibly including self-interactions). Technically, it amounts to replacing all $n$-point correlators by their Gaussian expression in terms of the two-point function. In the 2PI formalism this is typically obtained as the lowest nontrivial order of a loop expansion of the 2PI effective action. This neglects dissipation due to direct collisions and memory effects \cite{Berges:2004vw} and can be described by an effective bilinear Hamiltonian. 

Here, we consider the evolution equations for the equal-time two-point correlators of the (anti)neutrino system corresponding to a general bilinear Hamiltonian, to be specified later. We discuss the cases of Dirac and Majorana neutrino fields separately. 

\subsection{Dirac neutrinos}

We wish to obtain the evolution equations for the neutrino two-point correlators in the case where the effective Hamiltonian takes the general bilinear form  (we use $\hbar=c=1$ units)
\beq\label{e:Heff}
H_{\rm eff}(t) = \int d^3 {x}\, \bar{\psi}_{i}(t,\vec{x})\Gamma_{ij}(t,\vec{x}) \psi_{j}(t,\vec{x}),
\eeq
where $\psi_i$ denotes the $i$th component of the neutrino field in the mass basis.\footnote{We work here in the mass eigenstate basis, or propagation basis in vacuum. Alternatively one can write equivalent expressions using the matter basis, which is the basis that instanteneously diagonalizes the total neutrino Hamiltonian (including the neutrino-matter interaction and the self-interaction). Note that in the flavour basis the (anti)neutrino creation and annihilation operators do not satisfy the usual anticommutation relations, unless neutrinos are relativistic \cite{Giunti:1991cb}.} The explicit expression of the kernel $\Gamma$ is not needed here and will be specified later. At each time the spatial Fourier decomposition of a Dirac neutrino field reads
\begin{align}\label{e:field}
\psi_{i} (t,\vec{x}) =  \int_{\vec p,s} e^{i \vec p \cdot \vec{x}} \,\psi_{i} (t,\vec p,s),
\end{align}
with
\begin{align}\label{e:field2}
\psi_{i} (t,\vec p,s)= a_{i}(t,\vec p,s)u_{i}({\vec p,s}) + b_{i} ^{\dagger}(t,-\vec p,s) v_{i} (-{\vec p,s}),
\end{align}
where we note $\int_{\vec p}\equiv \int {d^3 {p} \over{(2 \pi)^3}}$ and $\int_{\vec p,s}\equiv \int_{\vec p}\sum_s $. Here, $a_i(t,\vec p,s)$ and $b_i(t,\vec p,s)$ are the standard particle and antiparticle annihilation operators (in the Heisenberg picture) for (anti)neutrinos of mass $m_i$, momentum $\vec p$, and helicity $s$. The nonzero equal-time anticommutation relations read
\beq\label{e:commutators1}
\{ a_{i}(t,\vec p,s), a^{\dagger}_{j}(t,\vec p\,',s') \} 
 = (2 \pi)^3 \delta^{(3)}(\vec p - \vec p\,')\delta_{ss'}\delta_{i j},
\eeq 
and similarly for the antiparticle operators. 
The Dirac spinors corresponding to mass eigenstates $i$ are normalized as (no sum over $i$)
\begin{equation}
 u^\dagger_i (\vec p,s)u_i(\vec p , s')=v^\dagger_i (\vec p,s)v_i(\vec p , s')=\delta_{ss'}.
\end{equation} 
In the flavor basis, the field operator  (possibly including sterile neutrinos) is obtained as 
\beq
 \psi_\alpha(t,\vec x)=U_{\alpha i}\,\psi_i(t,\vec x),
\eeq
where $U$ is the Maki-Nakagawa-Sakata-Pontecorvo unitary matrix \cite{Maki:1962mu}. In the three-flavor framework, the corresponding neutrino mixing angles are precisely measured, while the Dirac and two-Majorana CP-violating phases are still unknown \cite{Giunti:2007ry}. 

In the case of Dirac neutrinos, we neglect possible lepton-number-violating correlators. The set of equal-time two-point correlators is fully characterized by
\begin{align}\label{e:rho}
\rho_{ij}(t,\vec q,h,\vec q\,'\!,h')  &= \langle a^{\dagger}_{j}(t,\vec q\,'\!,h') a_{i}  (t,\vec q,h)  \rangle,\\
\label{e:arho}
\bar{\rho}_{ij}(t,\vec q,h,\vec q\,'\!,h') &= \langle b^{\dagger}_{i}(t,\vec q,h)   b_{j} (t,\vec q\,'\!,h')  \rangle,\\ 
\label{e:kappa}
\kappa_{ij}(t,\vec q,h,\vec q\,'\!,h') &= \langle b_{j}(t,\vec q\,'\!,h')   a_{i} (t,\vec q,h)  \rangle, \\
\label{e:kappastar}
\kappa^\dagger_{ij}(t,\vec q,h,\vec q\,'\!,h') &= \langle a^{\dagger}_{j} (t,\vec q\,'\!,h') b^{\dagger}_{i}(t,\vec q,h) \rangle,
\end{align}
where the brackets denote the quantum and statistical average over the medium through which the neutrinos are propagating. Here, $\rho$ and $\bar\rho$ describe generalized particle and antiparticle number densities\footnote{For $\bar{\rho}$, we employ the same convention as in Ref.~\cite{Sigl:1992fn}, and not the one adopted in Ref.~\cite{Volpe:2013uxl}.}, whereas $\kappa$ and $\kappa^\dagger$ correspond to particle-antiparticle pair correlations. Note that the particle and antiparticle correlators, $\rho$ and $\bar\rho$, include normal densities with all possible helicity states, including the ``wrong'' ones, e.g., $\rho_{ij}(t,\vec q,+,\vec q\,',+)$ or $\bar \rho_{ij}(t,\vec q,-,\vec q\,',-)$, as well as possible coherence terms such as $\rho_{ij}(t,\vec q,-,\vec q\,',+)$. Our notation is such that the Hermitian conjugation in \Eqn{e:kappastar} involves both mass indices and space-time variables, i.e., 
\begin{equation}
\kappa^\dagger_{ij}(t,\vec q,h,\vec q\,'\!,h') = \kappa^*_{ji}(t,\vec q\,'\!,h'\!,\vec q,h) .
\end{equation}

Let us derive the general evolution equations for the two-point correlators \eqn{e:rho}--\eqn{e:kappastar}. To simplify the derivation and presentation, it is useful to introduce some notations. We first define the spinor products
\begin{align}
\label{e:g1}
&\Gamma_{ij}^{\nu\nu}(t,\vec q,h,{\vec q\,'\!,h'}) = \bar{u}_{i}(\vec q,h)\tilde\Gamma_{ij}(t,\vec q-\vec q\,')u_{j}({\vec q\,'\!,h'}) ,  \\
\label{e:g2}
&\Gamma_{ij}^{\bar\nu\bar\nu}(t,\vec q,h,\vec q\,'\!,h') = \bar{v}_{i}(\vec q,h)\tilde\Gamma_{ij}(t,-\vec q+\vec q\,')v_{j}({\vec q\,'\!,h'}),
\end{align}
and
\begin{align}
\label{e:g3}
&\hspace*{-.08cm}\Gamma_{ij}^{\nu\bar\nu}(t,\vec q,h,\vec q\,'\!,h') = \bar{u}_{i}(\vec q,h)\tilde\Gamma_{ij}(t,\vec q+\vec q\,')v_{j}({\vec q\,'\!,h'}) ,  \\
\label{e:g4}
&\hspace*{-.08cm}\Gamma_{ij}^{\bar\nu\nu}(t,\vec q,h,\vec q\,'\!,h') = \bar{v}_{i}(\vec q,h)\tilde\Gamma_{ij}(t,-\vec q-\vec q\,')u_{j}({\vec q\,'\!,h'}),
\end{align}
with the Fourier transform of the mean-field defined as
 \begin{align}\label{e:fourier}
 \Gamma_{ij}(t,\vec{x}) = \int_{\vec p}  e^{i \vec p \cdot \vec{x}}\, \tilde\Gamma_{ij}(t,\vec p\,).
  \end{align}
The effective Hamiltonian \eqn{e:Heff} can then be written as
\begin{align}\label{e:momentum}
H_{\rm eff}(t)  =\int_{\vec p,s,\vec p'\!,s'}\!\!\Big[&\Gamma_{ij}^{\nu\nu}(t,\vec p,s,\vec p\,'\!,s')\, a^{\dagger}_{i} (t,\vec p,s)a_{j} (t,\vec p\,'\!,s')  \nonumber  \\ 
 + &   \Gamma_{ij}^{\bar\nu\bar\nu}(t,\vec p,s,\vec p\,'\!,s') \,b_{i}(t,\vec p,s)   b^{\dagger}_{j}(t,\vec p\,'\!,s')   \nonumber    \\
 + &   \Gamma_{ij}^{\nu\bar\nu}(t,\vec p,s,\vec p\,'\!,s')   \, a^{\dagger}_{i}(t,\vec p,s)  b^{\dagger}_{j} (t,\vec p\,'\!,s')  \nonumber    \\ 
+ &   \Gamma_{ij}^{\bar\nu\nu}(t,\vec p,s,\vec p\,'\!,s')   \, b_{i}(t,\vec p,s)  a_{j} (t,\vec p\,'\!,s')   \Big] .
 \end{align}
We use a matrix notation in both mass indices and space-time variables in which, for instance, $\Gamma_{ij}^{\nu\nu}(t,\vec q,h,{\vec q\,',h'})$ is the matrix element of $\Gamma^{\nu\nu}(t)$. The matrix product involves both a sum over the discrete (mass/flavor and helicity) indices and an integral over the continuous (momentum) variables, i.e.,
\begin{equation}
 [A\cdot B]_{ij}(\vec q, h,\vec q\,'\!,h')\equiv\int_{\vec p,s}A_{ik}(\vec q, h,\vec p,s)B_{kj}(\vec p, s,\vec q\,'\!, h'). 
\end{equation}
In this notation, the operators $a(t)$ and $b(t)$ are to be seen as column vectors and $\rho(t)$, $\bar\rho(t)$, and $\kappa(t)$ in Eqs.~\eqn{e:rho}--\eqn{e:kappastar} as well as the $\Gamma(t)$'s in Eqs.~\eqn{e:g1}--\eqn{e:g4} as matrices. The effective Hamiltonian \eqn{e:momentum} takes the compact form
\begin{align}\label{e:momentum-matrix}
H_{\rm eff}(t)  &=a^\dagger(t)\cdot\Gamma^{\nu\nu}(t)\cdot a (t)+b(t)\cdot\Gamma^{\bar\nu\bar\nu}(t)\cdot b^\dagger (t)  \nonumber  \\ 
  &+a^\dagger(t)\cdot\Gamma^{\nu\bar\nu}(t)\cdot b^\dagger (t)+b(t)\cdot\Gamma^{\bar\nu\nu}(t)\cdot a (t)  ,
 \end{align}
and the equal-time anticommutation relations \eqn{e:commutators1} read
\begin{equation}
\label{eq:anticmatrix}
\left\{a(t),a^\dagger(t)\right\}=\mathds{1}.
\end{equation}
The Hermiticity of the effective Hamiltonian implies 
\begin{align}
\left[\Gamma^{\nu\nu}(t)\right]^\dagger&=\Gamma^{\nu\nu}(t),\\
\left[\Gamma^{\bar\nu\bar\nu}(t)\right]^\dagger&=\Gamma^{\bar\nu\bar\nu}(t),\\
\label{eq:ksfdasdf}
\left[\Gamma^{\nu\bar\nu}(t)\right]^\dagger&=\Gamma^{\bar\nu\nu}(t).
\end{align}

The evolution equations for the correlator \eqn{e:rho} can be obtained as
\begin{align}\label{e:Ehrenrho}
i \dot{\rho}_{ij}(t,\vec q,h,\vec q\,'\!,h') = \langle  [a^{\dagger}_{j}(t,\vec q\,',h')a_{i}(t,\vec q,h), H_{\rm eff}(t) ] \rangle .
\end{align}
Using \Eqn{e:momentum} and the anticommutation relations \eqn{e:commutators1}, one easily expresses the right-hand side in terms of the two-point correlators \eqn{e:rho}--\eqn{e:kappastar}: 
\begin{align}\label{e:rhoev}
i \dot{\rho}_{ij}(t,\vec q,h,\vec q\,'\!,h')  = \! \int_{\vec p,s}\!\! \Big[&\Gamma_{ik}^{\nu\nu}(t,\vec q,h,\vec p,s)\, \rho_{k j}(t,\vec p,s,\vec q\,'\!,h')  \nonumber  \\ 
- & \rho_{i k}(t,\vec q,h,\vec p,s) \,  \Gamma_{kj}^{\nu\nu}(t,\vec p,s, \vec q\,'\!,h') \nonumber  \\ 
+ &  \Gamma_{ik}^{\nu\bar\nu}(t,\vec q,h,\vec p,s)  \,\kappa^\dagger_{kj}(t, \vec p,s,\vec q\,'\!,h')   \nonumber  \\ 
- & \kappa_{i k} (t,\vec q,h, \vec p,s)\,\Gamma_{k j}^{\bar\nu\nu}(t,\vec p,s,\vec q\,'\!,h') \Big]   .
\end{align}
Using the matrix notation introduced above, this takes the compact form
\begin{align}\label{e:rhoev}
i \dot{\rho}(t)  &=  \Gamma^{\nu\nu}(t)\cdot \rho(t) - \rho(t) \cdot \Gamma^{\nu\nu}(t) \nonumber  \\ 
&+  \Gamma^{\nu\bar\nu}(t)  \cdot \kappa^\dagger(t)  - \kappa (t)\cdot\Gamma^{\bar\nu\nu}(t).
\end{align}
The evolution equation for the antineutrino density \eqn{e:arho} is obtained along the same lines as
\begin{align}\label{e:arhoev}
i \dot{\bar \rho}(t)  &=  \Gamma^{\bar\nu\bar\nu}(t)\cdot\bar \rho(t) - \bar \rho(t) \cdot \Gamma^{\bar\nu\bar\nu}(t) \nonumber  \\ 
&-  \Gamma^{\bar\nu\nu}(t)  \cdot \kappa(t)  + \kappa^\dagger (t)\cdot\Gamma^{\nu\bar\nu}(t).
\end{align}
Finally, the evolution equation for the neutrino-antineutrino pairing correlator reads 
\begin{align}
i \dot{\kappa}(t)& = \Gamma^{\nu\nu}(t)\cdot {\kappa}(t) -  {\kappa}(t) \cdot \Gamma^{\bar\nu\bar\nu}(t)  \nonumber  \\ 
 & -  \Gamma^{\nu\bar\nu}(t) \cdot\bar{\rho}(t)- \rho(t)\cdot \Gamma^{\nu\bar\nu}(t)\nonumber\\
 \label{e:kev}
 &+  \Gamma^{\nu\bar\nu}(t) 
\end{align}
and similarly for $\kappa^\dagger(t)$.

We mention that these general evolution equations can be gathered together by introducing a further level of matrix notation. Following Ref.~\cite{Volpe:2013uxl}, we define
\begin{equation}\label{e:genH}
{\cal H} (t)= \left(
\begin{array}{cc}   
\Gamma^{\nu\nu}(t) & \Gamma^{\nu\bar\nu}(t) \\
\Gamma^{\bar\nu\nu}(t)  & \Gamma^{\bar\nu\bar\nu}(t)  \end{array}
\right)
\eeq
and
\begin{equation}\label{e:genR}
{\cal R}(t) =\left(
\begin{array}{cc}   
 \rho(t) &  \kappa (t) \\
\kappa^{\dagger}(t) & \mathds{1} - \bar{\rho}(t) \end{array}
\right),
\end{equation}
in terms of which Eqs. (\ref{e:rhoev})--(\ref{e:kev}) can be rewritten as
\begin{equation}\label{e:matrixform}
i\, \dot{\!{\cal R}} (t)= \left[ {\cal H}(t),{\cal R}(t)\right].
\end{equation}
Note that the conservation of the total lepton number, defined as
\beq
\label{eq:leptonnumber}
 L={\rm tr}\left[\rho(t)\!-\!\bar\rho(t)\right]=\!\!\int_{\vec p,s}\!\!\left[\rho_{ii}(t,\vec p,s,\vec p,s)-\bar\rho_{ii}(t,\vec p,s,\vec p,s)\right]\!,
\eeq
follows immediately from ${\rm tr}\,\,\dot{\!\cal R}(t)=0$.

We emphasize that the evolution equations \eqn{e:rhoev}--\eqn{e:kev} include all possible correlators between the various helicity states of both particles and antiparticles and encompass all mean-field evolution equations discussed so far in the literature for Dirac neutrinos. Let us now discuss  the case of Majorana neutrinos.

\subsection{Majorana neutrinos}
\label{sec:majorana}

The general mean-field Hamiltonian for Majorana fields reads\footnote{We use the same notation as in the Dirac case for the kernel $\Gamma$ for simplicity, although one should keep in mind that the respective vacuum (free) contributions differ by a factor of $1/2$ to account for the different number of degrees of freedom \cite{Giunti:2007ry}.}
\beq\label{e:MHeff}
H_{\rm eff}^M(t) = \int d^3 {x}\, \bar{\psi}_{i}^M(t,\vec{x})\Gamma_{ij}(t,\vec{x}) \psi_{j}^M(t,\vec{x}),
\eeq
where
\begin{align}\label{e:Majofield}
\psi_{i}^M (t,\vec{x}) =  \int_{\vec p,s} e^{i \vec p \cdot \vec{x}} \,\psi_{i}^M (t,\vec p,s),
\end{align}
with
\begin{align}\label{e:Majofield2}
\psi_{i}^M (t,\vec p,s)= a_{i}(t,\vec p,s)u_{i}({\vec p,s}) + a_{i}^{\dagger}(t,-\vec p,s) v_{i} (-{\vec p,s}).
\end{align}

The evolution equations for Majorana fields can be formally obtained from those derived in the previous section by replacing the antiparticle operators by particle ones: $b\to a$. Using the notation introduced above, \Eqn{e:MHeff} can be written in the symmetric form
\begin{align}\label{e:momentum-matrix-bis}
H_{\rm eff}^M(t)  =\frac{1}{2}\Big[&a^\dagger(t)\cdot\Gamma^{\nu\nu}_M(t)\cdot a (t)+a(t)\cdot\Gamma^{\bar\nu\bar\nu}_M(t)\cdot a^\dagger (t)  \nonumber  \\ 
  +&a^\dagger(t)\cdot\Gamma^{\nu\bar\nu}_M(t)\cdot a^\dagger (t)+a(t)\cdot\Gamma^{\bar\nu\nu}_M(t)\cdot a (t)\Big] ,
 \end{align}
where we defined 
\begin{align}
\label{eqn:MajoHam1}
 \Gamma^{\nu\nu}_M(t)&=\Gamma^{\nu\nu}(t)-[\Gamma^{\bar\nu\bar\nu}(t)]^T ,\\
\label{eqn:MajoHam2}
 \Gamma^{\bar\nu\bar\nu}_M(t)&=\Gamma^{\bar\nu\bar\nu}(t)-[\Gamma^{\nu\nu}(t)]^T, \\
\label{eqn:MajoHam3}
 \Gamma^{\nu\bar\nu}_M(t)&=\Gamma^{\nu\bar\nu}(t)-[\Gamma^{\nu\bar\nu}(t)]^T,\\
\label{eqn:MajoHam4}
 \Gamma^{\bar\nu\nu}_M(t)&=\Gamma^{\bar\nu\nu}(t)-[\Gamma^{\bar\nu\nu}(t)]^T.
 \end{align}
Note the relations
\begin{align}
 \label{eq:Mrel}
 \left[\Gamma^{\nu\nu}_M(t)\right]^T&=-\Gamma^{\bar\nu\bar\nu}_M(t),\\
 \left[\Gamma^{\nu\bar\nu}_M(t)\right]^T&=-\Gamma^{\nu\bar\nu}_M(t),\\
 \label{eq:Mrelbis}
 \left[\Gamma^{\bar\nu\nu}_M(t)\right]^T&=-\Gamma^{\bar\nu\nu}_M(t),
 \end{align}
where the superscript $T$ stands for transposition in both mass indices and space-time variables.

The most general set of equal-time two-point correlators include generalized particle densities $\rho_M=\langle a^\dagger a\rangle$ and pair correlations $\kappa_M=\langle aa \rangle$. We define the latter as in Eqs.~\eqn{e:rho}--\eqn{e:kappastar}, with the obvious relations 
\beq
\label{eq:MM}
 \bar \rho_M(t)=\Big[\rho_M(t)\Big]^T\quad{\rm  and} \quad \kappa_M(t)=-\Big[\kappa_M(t)\Big]^T.
\eeq
Although Majorana particles are their own antiparticles, it is common practice to refer to negative- and positive-helicity states as ``particles'' and ``antiparticles''. As in the Dirac case, the particle and pair correlators $\rho_M$ and $\kappa_M$ include all possible helicity states. In particular, both the usual ``particle'' and ``antiparticle'' densities are encoded in $\rho_M$, as $\rho^M_{ij}(t,\vec q,-,\vec q\,',-)$ and $\rho^M_{ij}(t,\vec q,+,\vec q\,',+)$. Equivalently, the latter is given by $\bar\rho^M_{ji}(t,\vec q\,',+,\vec q,+)$, using \Eqn{eq:MM}. The other, nondiagonal helicity components of $\rho_M$ describe ``particle-antiparticle'' coherence of the type discussed in Refs.~\cite{Dvornikov:2011dv,deGouvea:2012hg,Vlasenko:2013fja}. 

The pair correlations encoded in $\kappa_M$ describe other kinds of coherences of either the ``particle-antiparticle'' type, with $\kappa^M_{ij}(t,\vec q,+,\vec q\,',-)$, or ``particle-particle'' and ``antiparticle-antiparticle'' type, with $\kappa^M_{ij}(t,\vec q,-,\vec q\,',-)$ and $\kappa^M_{ij}(t,\vec q,+,\vec q\,',+)$ respectively, as was first discussed in Ref.~\cite{Volpe:2013uxl}. Note that, in contrast to the Dirac case, the Majorana pair correlations $\kappa_M$ violate the total lepton number.\footnote{There is no contradiction with the total lepton number conservation discussed in \Eqn{eq:leptonnumber}, since in the Majorana case, ${\rm tr}[\rho_M(t)-\bar\rho_M(t)]={\rm tr}[\rho_M(t)-\rho^T_M(t)]=0$ is not to be intrerpreted as the lepton number. We have, instead $L_M(t)=\tr\,\rho_M(t)$, which is clearly not conserved in the presence of the lepton-number violating pair correlations~$\kappa^{}_M$:
$i\dot L^{}_M(t)=\tr[\Gamma_M^{\nu\bar\nu}(t)  \cdot \kappa_M^\dagger(t)  - \kappa^{}_M (t)\cdot\Gamma_M^{\bar\nu\nu}(t)]$.}

The general mean-field equations \eqn{e:rhoev} and \eqn{e:kev} derived in the previous section for Dirac fields keep the very same form for Majorana fields with the replacements $\Gamma\to\Gamma_M$, $\rho\to\rho_M$, and $\kappa\to\kappa_M$. Specifically, by introducing 
\begin{equation}\label{e:genHMajo}
{\cal H}_M (t)= \left(
\begin{array}{cc}   
\Gamma_M^{\nu\nu}(t) & \Gamma_M^{\nu\bar\nu}(t) \\
\vspace*{-.35cm}\\
\Gamma_M^{\bar\nu\nu}(t) & -\left[\Gamma_M^{\nu\nu}(t)\right]^T \end{array}
\right)
\eeq
and
\begin{equation}\label{e:genRMajo}
{\cal R}_M(t) =\left(
\begin{array}{cc}   
 \rho_M(t) &  \kappa_M (t) \\
\kappa_M^{\dagger}(t) & \mathds{1} - {\rho}_M^T(t) \end{array}
\right),
\end{equation}
one has
\begin{equation}\label{e:matrixformMajo}
i\, \dot{\!{\cal R}}_M (t)= \left[ {\cal H}_M(t),{\cal R}_M(t)\right].
\end{equation}
One easily checks that the Dirac evolution equations in the Dirac case reduce to those in the Majorana case when the symmetry properties \eqn{eq:Mrel}--\eqn{eq:MM} are imposed. As a consequence, the observation we made in the Dirac case holds here too:  nontrivial pair correlations $\kappa_M$ may develop and  backreact on the evolution of the normal densities $\rho_M$ whenever $\Gamma^{\nu\bar\nu}_M\neq0$.

\section{Pairing correlations}
\label{sec:pair}

We now focus on particle-antiparticle mixing resulting from possible nonzero pairing correlations \cite{Volpe:2013uxl}. These are discarded in most existing treatments of in-medium neutrino propagation (see, e.g., Refs.~\cite{Sigl:1992fn,Vlasenko:2013fja}), based on the argument that, in the free theory, they undergo rapid oscillations on a time scale $\sim 1/[\epsilon_{i,q}+\epsilon_{j,q}]$, where $\epsilon_{i,q}=(q^2+m_i^2)^{1/2}$ is the (anti)particle energy, and thus they average to zero on the typical time scales of interest for, e.g., neutrino flavor conversion. As we shall see below, this argument must be reconsidered in an interacting theory since such pairs generally receive contributions from the (anti)neutrino densities $\rho$ and $\bar\rho$. In particular, this occurs when the medium with which neutrinos interact is inhomogeneous and/or anisotropic. 

In the following, we consider the case of a spatially homogeneous system and we take the ultrarelativistic (massless) limit for neutrinos since the role of such correlations is not controlled by the neutrino mass. Nonrelativistic corrections are discussed in Appendix \ref{appsec:total}. We keep working in the mass basis although in this limit the evolution equations are equally valid in the flavor basis since the spinors appearing in Eqs.~\eqn{e:g1}--\eqn{e:g4} do not depend on the mass/flavor index. We first discuss the Dirac case and give the necessary modifications for the Majorana case.

\subsection{Evolution equations for homogeneous systems in the ultrarelativistic limit}

In the ultrarelativistic limit, and assuming Standard Model $V$-$A$ interactions, it is sufficient to restrict oneself to the subset of two-point correlators in Eqs.~\eqn{e:rho}--\eqn{e:kappastar} involving negative-helicity particle or positive-helicity antiparticle operators. For spatially homogeneous systems these read
\begin{align}\label{e:homogrho}
\rho_{ij}(t,\vec q,-,\vec q\,'\!,-)  &= (2 \pi)^3 \delta^{(3)}(\vec q - \vec q\,') \rho_{ij}(t,\vec q\,),\\
\label{e:homogrhobar}
\bar \rho_{ij}(t,\vec q,+,\vec q\,'\!,+)  &= (2 \pi)^3 \delta^{(3)}(\vec q - \vec q\,') \bar\rho_{ij}(t,-\vec q\,),\\
\label{e:homogk}
\kappa_{ij}(t,\vec q,-,\vec q\,'\!,+)  &= (2 \pi)^3  \delta^{(3)}(\vec q + \vec q\,') \kappa_{ij}(t,\vec q\,),\\
\label{e:homogkdag}
\kappa^\dagger_{ij}(t,\vec q,+,\vec q\,'\!,-) & = (2 \pi)^3  \delta^{(3)}(\vec q + \vec q\,') \kappa^\dagger_{ij}(t,-\vec q\,).
\end{align}
For later convenience, we define the density of particles and antiparticles with momentum $\vec q$ as $\rho_{ij}(t,\vec q\,)$ and $\bar\rho_{ij}(t,-\vec q\,)$, respectively. Note also that \Eqn{e:homogkdag} ensures that $\kappa^\dagger_{ij}(t,\vec q\,)=\kappa^*_{ji}(t,\vec q\,)$. Spatial homogeneity also implies
\begin{equation}
\tilde\Gamma_{ij}(t,\vec q\,) = (2 \pi)^3 \delta^{(3)}(\vec q\,) \,\tilde{\Gamma}_{ij}(t),
\end{equation}
and we define, in accordance with Eqs.~\eqn{e:homogrho}--\eqn{e:homogkdag},
\begin{align}
\label{e:g1hom}
\Gamma_{ij}^{\nu\nu}(t,\vec q,-,\vec q\,'\!,-) & =  (2 \pi)^3 \delta^{(3)}(\vec q - \vec q\,') \Gamma^{\nu\nu}_{ij}(t,\vec q\,),\\
\label{e:g2hom}
\Gamma_{ij}^{\bar\nu\bar\nu}(t,\vec q,+,\vec q\,'\!,+) & = (2 \pi)^3 \delta^{(3)}(\vec q - \vec q\,')\Gamma^{\bar\nu\bar\nu}_{ij}(t,-\vec q\,),\\
\label{e:g3hom}
\Gamma_{ij}^{\nu\bar\nu}(t,\vec q,-,\vec q\,'\!,+) & =  (2 \pi)^3 \delta^{(3)}(\vec q + \vec q\,')\Gamma^{\nu\bar\nu}_{ij}(t,\vec q\,),\\
\label{e:g4hom}
\Gamma_{ij}^{\bar\nu\nu}(t,\vec q,+,\vec q\,'\!,-) & =  (2 \pi)^3 \delta^{(3)}(\vec q + \vec q\,')\Gamma^{\bar\nu\nu}_{ij}(t,-\vec q\,).
 \end{align}
As emphasized above, the Dirac spinors do not depend on the mass index in the ultrarelativistic limit. Writing $u(\vec p\,)\equiv u_i(\vec p,-)$ and $v(\vec p\,)\equiv v_i(\vec p,+)$, we have
\begin{align}
\label{eq:diag1}
\Gamma^{\nu\nu}_{ij}(t,\vec q\,)&=\bar u(\vec q\,)\tilde\Gamma_{ij}(t)u(\vec q\,),\\
\label{eq:diag2}
\Gamma^{\bar\nu\bar\nu}_{ij}(t,\vec q\,)&=\bar v(-\vec q\,)\tilde\Gamma_{ij}(t)v(-\vec q\,),\\
\label{eq:nondiag1}
\Gamma^{\nu\bar\nu}_{ij}(t,\vec q\,)&=\bar u(\vec q\,)\tilde\Gamma_{ij}(t)v(-\vec q\,),\\
\label{eq:nondiag2}
\Gamma^{\bar\nu\nu}_{ij}(t,\vec q\,)&=\bar v(-\vec q\,)\tilde\Gamma_{ij}(t)u(\vec q\,).
\end{align}

The evolution equations \eqn{e:rhoev}--\eqn{e:kev} reduce to
\begin{align}
i \dot{\rho}(t,\vec q\,) & = \left[{\Gamma}^{\nu\nu}(t,\vec q\,), \rho(t,\vec q\,)\right] \nonumber  \\ 
\label{e:rhoevhom}
 &+   {\Gamma}^{\nu\bar\nu} (t,\vec q\,)\cdot \kappa^\dagger(t,\vec q\,) -  \kappa (t,\vec q\,) \cdot {\Gamma}^{\bar\nu\nu} (t,\vec q\,),
\end{align}
\begin{align}
i \dot{\bar \rho}(t,\vec q\,)  &=  \left[\Gamma^{\bar\nu\bar\nu}(t,\vec q\,),\bar \rho(t,\vec q\,)\right] \nonumber  \\ 
\label{e:arhoevhom}
&-  \Gamma^{\bar\nu\nu}(t,\vec q\,)  \cdot \kappa(t,\vec q\,)  + \kappa^\dagger (t,\vec q\,)\cdot\Gamma^{\nu\bar\nu}(t,\vec q\,),
\end{align}
and
\begin{align}
i \dot{\kappa}(t,\vec q\,)& = \Gamma^{\nu\nu}(t,\vec q\,)\cdot {\kappa}(t,\vec q\,) -  {\kappa}(t,\vec q\,) \cdot \Gamma^{\bar\nu\bar\nu}(t,\vec q\,)  \nonumber  \\ 
 & -  \Gamma^{\nu\bar\nu}(t,\vec q\,) \cdot\bar{\rho}(t,\vec q\,)- \rho(t,\vec q\,)\cdot \Gamma^{\nu\bar\nu}(t,\vec q\,)\nonumber\\
\label{e:kevhom}
 &+  \Gamma^{\nu\bar\nu}(t,\vec q\,) ,
\end{align}
where the dot now involves a simple matrix product in mass/flavor indices, which we have left implicit here. Our convention for the signs of momenta in Eqs.~\eqn{e:homogrho}--\eqn{e:g4hom} is such that all quantities appearing in the above evolution equations involve the same momentum $\vec q$. 

These equations can be conveniently rewritten in terms of the $2N_f\times2N_f$ matrices 
\begin{equation}\label{e:genH2}
{\cal H}(t,\vec q\,) = \left(
\begin{array}{cc}   
\Gamma^{\nu\nu}(t,\vec q\,) & \Gamma^{\nu\bar\nu}(t,\vec q\,) \\
\Gamma^{\bar\nu\nu}(t,\vec q\,)  & \Gamma^{\bar\nu\bar\nu} (t,\vec q\,) \end{array}
\right)
\end{equation}
and
\begin{equation}\label{e:genR2}
{\cal R}(t,\vec q\,) = \left(
\begin{array}{cc}   
 \rho(t,\vec q\,) &  \kappa(t,\vec q\,)  \\
\kappa^{\dagger}(t,\vec q\,) & \mathds{1} - \bar{\rho}(t,\vec q\,) \end{array}
\right)
\end{equation}
as\footnote{Note that for spatially homogeneous systems in the mean-field approximation there are infinitely many conserved quantities, one for each Fourier mode, as follows from ${\rm tr}\,\dot{\cal R}(t,\vec q\,)=0$, which implies that  $\rho_{ii}(t,\vec q\,)-\bar\rho_{ii}(t,\vec q\,)={\rm const}$. Such conservations laws are a particular feature of the mean-field dynamics, due to the neglect of momentum-changing collision processes and are not present in the exact theory. Note that they do not correspond to individual lepton number in each mode since they involve $\bar \rho_{ii}(t,\vec q\,)$, the density of antiparticles with momentum $-\vec q$, instead of $\bar \rho_{ii}(t,-\vec q\,)$. However, these conservation laws are, of course, compatible with total lepton number conservation since they imply
$
 \int_{\vec p} \left[\rho_{ii}(t,\vec p\,)-\bar\rho_{ii}(t,-\vec p\,)\right]={\rm const.}
$
}
\begin{equation}\label{e:matrixform2}
i\, \dot{\!{\cal R}}(t,\vec q\,) = [ {\cal H}(t,\vec q\,),{\cal R}(t,\vec q\,)].
\end{equation}

The corresponding evolution equations for Majorana neutrinos can be readily obtained from the previous ones following the discussion of Sec.~\ref{sec:majorana}. In the ultrarelativistic limit considered here, with $V$-$A$ interactions, one easily checks that the components of the extended mean-field Hamiltonians  for Dirac and Majorana neutrinos coincide: ${\cal H}_M(t,\vec q\,)={\cal H}(t,\vec q\,)$. The evolution equations for Majorana neutrinos thus take the very same form as above with the replacements $\rho(t,\vec q\,)\to\rho_M(t,\vec q\,)$,  $\bar\rho(t,\vec q\,)\to\bar\rho_M(t,\vec q\,)$, and $\kappa(t,\vec q\,)\to\kappa_M(t,\vec q\,)$. Here, $\rho_M$ describes the density of ``particles'' (negative-helicity states), $\bar\rho_M$ the density of ``antiparticles'' (positive-helicity states), and $\kappa_M$ the ``particle-antiparticle'' pair correlations. 

The first lines of both \Eqn{e:rhoevhom} and \Eqn{e:arhoevhom} correspond to those usually considered in most existing treatments of in-medium neutrino propagation, which neglect the pair correlator $\kappa$. However, from \Eqn{e:kevhom} one observes that the latter is sourced both by the particle and antiparticle densities $\rho$ and $\bar\rho$ and by the off-diagonal component $\Gamma^{\nu\bar\nu}$ of the mean-field Hamiltonian \eqn{e:genH2}. One should thus reconsider the usual argument described previously that neglects the pair correlations since these source terms have no reason to be rapidly oscillating. If the particle-antiparticle coupling $\Gamma^{\nu\bar\nu}$ is nonzero, pair correlations can be generated through \Eqn{e:kevhom} and have a nontrivial backreaction on the time evolution of the normal densities $\rho$ and $\bar\rho$. We now discuss the explicit form of the mean-field Hamiltonian \eqn{e:matrixform2} and the conditions for such a nontrivial  particle-antiparticle mixing term.

\subsection{Mean-field Hamiltonian in a dense anisotropic environment}
\label{sec:MF}

We consider a typical astrophysical environment, e.g., a core-collapse supernova, with a dense gas of self-interacting (anti)neutrinos in a background of electrons and nucleons.\footnote{One can generalize the present treatment to take into account possible external (e.g., magnetic) fields, other type of particles in the background, and/or nonstandard interactions.} For the sake of the argument, we assume (local) spatial homogeneity, although this idealization should be relaxed for simulations in realistic geometries. We assume an unpolarized and electrically neutral but otherwise possibly anisotropic matter background while we treat the relativistic neutrino gas in full generality. Matter anisotropies are important in multidimensional {supernovae} simulations \cite{Janka:2012}. Neutrino anisotropies have been shown to potentially play an important role in neutrino flavor conversion \cite{Duan:2010bg}.

A new ingredient of the present discussion is that, contrary to previous treatments, we keep track of the neutrino-antineutrino pair correlations in the expression of the mean-field Hamiltonian [\Eqn{e:Heff} or \Eqn{e:MHeff}]. To this aim we consider a regime where an extended mean-field approximation is still applicable and explicit collisions are neglected. Strictly speaking, in the typical situation of interest here---where the neutrino opacity is non-negligible---one should also include collisions. However, the mean-field approximation is simple enough to be manageable, and it allows us to pinpoint the potential role of neutrino-antineutrino pair correlations and the necessary ingredients for their presence.

\subsubsection{Vacuum and matter contributions}

Let us first consider the contributions to the kernel $\Gamma_{ij}(t,\vec x)$ from the free Hamiltonian, i.e.,  $(-i\vec \gamma\cdot\vec\nabla+m_i)\delta_{ij}$, and from the neutral- and charged-current interactions with the matter background. The former does not lead to any off-diagonal particle-antiparticle mixing term in \Eqn{e:genH2}: $\Gamma^{\nu\bar\nu}_{{\rm vac}}(t,\vec q\,)=0$ and
\beq
\label{eq:vacuumham}
 \Gamma^{\nu\nu}_{{\rm vac}}(t,\vec q\,)=-\Gamma^{\bar\nu\bar\nu}_{{\rm vac}}(t,\vec q\,)=h^0(q),
\eeq
 with $h^0(q)={\rm diag}(\epsilon_{i,q})$ in the mass eigenstate basis.

The matter contribution only concerns active neutrino species and is diagonal in the flavor basis. The neutral-current interactions with the electron, proton, and neutron backgrounds are flavor insensitive and give the mean-field kernel 
\begin{align}
 \Gamma^{\rm mat, NC}_{\alpha\beta}(t,\vec x)&={G_F \over{\sqrt{2}}}\delta_{\alpha\beta}\gamma^{\mu}(1-\gamma_5)\nn
 \label{eq:NCmat}
 &\times \sum_{f=e,p,n}\langle  \bar{\phi}_{f}(t,\vec{x})\gamma_{\mu}(c_V^f-c_A^f\gamma_5) \phi_{f}(t,\vec{x}) \rangle ,
\end{align}
where $\phi_f$ represents the field associated to the particle $f$ and where $c_V^f$ and $c_A^f$ are the corresponding vector and axial coupling constants. 
Assuming an unpolarized background with no antiparticles, the medium average in \Eqn{eq:NCmat} is easily computed as \cite{Giunti:2007ry}
\beq
 \tilde\Gamma^{\rm mat, NC}_{\alpha\beta}(t)={G_F \over{\sqrt{2}}}\delta_{\alpha\beta}\gamma^{\mu}(1-\gamma_5)\!\!\sum_{f=e,p,n}c_V^fJ^f_\mu(t),
\eeq
where we defined the four-velocity density for the particle of type $f$ as
\beq
 J^\mu_f(t)=2\int_{\vec p}v^\mu_f\rho_f(t,\vec p\,),
\eeq
with $v^\mu_f=p^\mu/E^f_p$ and $E_p^f=\sqrt{p^2+m_f^2}$. The assumption of vanishing electric charge current, i.e., $J_\mu^e(t)+J_\mu^p(t)=0$, guarantees that the contribution from protons and electrons to the neutral-current mean-field Hamiltonian cancel each other since $c_V^e=-c_V^p$.

Finally, the charged-current contribution from electrons is of the same form as \Eqn{eq:NCmat} with $c_V=c_A=1$ and yields
\beq
 \tilde\Gamma^{\rm mat, CC}_{\alpha\beta}(t)={G_F \over{\sqrt{2}}}\delta_{\alpha \beta}\delta_{\alpha e}\gamma^{\mu}(1-\gamma_5)J^e_\mu(t).
\eeq
Using $c_A^n=-1/2$, the total contribution from the matter background thus reads, in the flavor basis,
\beq
\label{eq:matterham}
 \tilde\Gamma^{\rm mat}_{\alpha\beta}(t)={G_F \over{\sqrt{2}}}\delta_{\alpha\beta}\gamma^{\mu}(1-\gamma_5)\left[J^e_\mu(t)\delta_{\alpha e}-\frac{1}{2}J_\mu^n(t)\right].
\eeq

\subsubsection{Neutrino self-interactions}

The neutrino self-interaction mean-field Hamiltonian can be obtained from the Standard Model neutral-current interaction term in the contact interaction approximation,
\begin{align}
\label{eq:hnc}
 H_{\rm int}^{\rm self}= {G_F \over{4\sqrt{2}}}\sum_{\alpha,\beta}\int d^3 {x}\,j_\alpha^\mu(t,\vec x) j_{\beta,\mu}(t,\vec x),
\end{align}
where $j_\alpha^\mu (t,\vec x)=\bar{\psi}_{\alpha}(t,\vec{x})\gamma^{\mu}(1-\gamma_5) \psi_{\alpha}(t,\vec{x})$. Note that, thanks to the flavor sums in \Eqn{eq:hnc}, the neutral-current Hamiltonian takes the very same form in the mass eigenstate basis. This is useful when one is interested in keeping track of the neutrino masses in the mean-field evolution equations; see Sec. \ref{sec:hel} and Appendix \ref{appsec:total}. Although this is unimportant in the present case, where we take the massless limit, we keep working in the mass eigenstate basis for notational consistency throughout the paper. The mean-field approximation to \Eqn{eq:hnc} can be simply obtained by replacing quadrilinear products of field operators with bilinear ones using a Wick-like formula and using the Fierz theorem to reorganize spinorial products \cite{Samuel:1993uw}. This leads to an expression of the form \eqn{e:Heff} with, in the mass basis,\footnote{The relation to the flavor basis is given by $ \Gamma_{ij} = U^\dagger_{i \alpha}\Gamma_{\alpha\beta}U_{\beta j}$.}
\begin{align}
\label{eq:mmm}
\Gamma^{\rm self}_{ij}(t,\vec x)&={G_F \over{\sqrt{2}}}\gamma_{\mu}(1-\gamma_5) \Big[T^\mu_{ij}(t,\vec x)+\delta_{ij}T^\mu_{kk}(t,\vec x)\Big],
\end{align}
where 
\beq
\label{eq:mmmm}
 T^\mu_{ij}(t,\vec x)=\frac{1}{2}\langle  \bar{\psi}_{j}(t,\vec{x})\gamma^{\mu}(1-\gamma_5) \psi_{i}(t,\vec{x}) \rangle.
\eeq

The mean-field kernel \eqn{eq:mmm} is easily expressed in terms of the (anti)particle densities and pair correlations \eqn{e:rho}-\eqn{e:kappastar}. In the homogenous case, the relevant spinor products are
\beq\label{e:spi1}
\bar{u}(\vec p\,)  \gamma^{\mu}(1-\gamma_5)  u({\vec p}) = \bar{v}(\vec p\,) \gamma^{\mu}(1-\gamma_5) v({\vec p}) = 2n^\mu(\hat p)
\eeq
and
\begin{align}\label{e:spi2} 
\bar{v}(-\vec p\,)  \gamma^\mu(1-\gamma_5) u({\vec p}) &= 2\epsilon^\mu(\hat p).
\end{align}
Here,  we have introduced the light-like four-vectors 
\beq
\label{eq:lightlike}
 n^\mu(\hat p)=\left(\begin{tabular}{c}$1$\\$\hat p$\end{tabular}\right)\quad{\rm  and}\quad \epsilon^\mu(\hat p)=\left(\begin{tabular}{c}$0$\\$\hat\epsilon_p$\end{tabular}\right),
\eeq
where $\hat p=\vec p/ p$ denotes the unit vector in the direction of $\vec p$ and the pair of complex vectors $(\hat\epsilon_p,\hat\epsilon^*_p)$ spans the plane orthogonal to $\vec p$, with\footnote{In terms of an oriented triad of real orthogonal unit vectors $(\hat p,\hat p_\theta, \hat p_\phi)$---for instance the standard unit vectors associated to $\vec p$ in spherical coordinates---one has $\hat\epsilon_p=\hat p_\theta-i\hat p_\phi$. }  $\hat\epsilon_p \cdot\hat\epsilon_p=0$, $\hat\epsilon_p\cdot \hat\epsilon_p^*=2$.  The four-vectors \eqn{eq:lightlike} satisfy 
\beq
 n^\mu(\hat p) n_\mu(\hat p)=n^\mu(\hat p) \epsilon_\mu(\hat p)=\epsilon^\mu(\hat p)\epsilon_\mu(\hat p)=0
\eeq 
and 
\beq
 \epsilon^\mu(\hat p)\epsilon^*_\mu(\hat p)=-2.
\eeq 
Note also that $\epsilon_\mu(-\hat p)=\epsilon_\mu^*(\hat p)$. We thus obtain, for the kernel \eqn{eq:mmmm}, leaving the mass/flavor indices implicit,\footnote{We use the fact that, in the vacuum, $\langle0|\bar\psi\gamma^\mu(1-\gamma_5)\psi|0\rangle=\int_{\vec p}\bar{v}(\vec p\,) \gamma^{\mu}(1-\gamma_5) v(\vec p\,)=0$ by Lorentz symmetry (this requires a Lorentz-invariant ultraviolet regulator).}
\beq
 \label{e:gmf}
 \tilde\Gamma^{\rm self}(t)={G_F \over{\sqrt{2}}}\gamma^{\mu}(1-\gamma_5) \Big[T_\mu(t)+\mathds{1}\,\tr\,T_\mu(t)\Big],
\eeq
with
\beq 
\label{e:gmfbis}
 T_\mu (t)=  \int_{\vec p} \Big\{ n_\mu(\hat p)\ell (t,{\vec p}) + \epsilon_\mu(\hat p) \kappa(t,{\vec p})  +  \epsilon_\mu^*(\hat p) \kappa^\dagger(t,{\vec p}) \Big\},
\eeq
where we defined 
\beq
\label{eq:net}
 \ell(t,\vec q\,)=\rho (t,{\vec q}) - \bar{\rho}(t,-{\vec q}).
\eeq

\subsubsection{Total mean-field Hamiltonian}

The components \eqn{eq:diag1}--\eqn{eq:nondiag2} of the total mean-field Hamiltonian \eqn{e:genH2} are readily obtained using Eqs.~\eqn{e:spi1} and \eqn{e:spi2}. 
Collecting the contributions from the previous subsections, we finally get
\begin{align}
\label{eq:llmml}
 \Gamma^{\nu\nu}(t,\vec q\,)&= S(t, q)-\hat q\cdot \vec V(t),\\
 \Gamma^{\bar\nu\bar\nu}(t,\vec q\,)&=\bar S(t, q)+\hat q\cdot \vec V(t),\\
 \Gamma^{\nu\bar\nu}(t,\vec q\,)&= -\hat\epsilon_q^*\cdot\vec V(t),
\end{align}
with the $N_f\times N_f$ scalar and vector matrices
\begin{align}
\label{eq:scalar}
 S(t, q)&=h^0(q)+h^{\rm mat}(t)+\sqrt{2}G_F\Big[T_0(t)+\mathds{1}\,\tr\,T_0(t)\Big],\\
\label{eq:scalarbar}
 \bar S(t, q)&=-h^0(q)+h^{\rm mat}(t)+\sqrt{2}G_F\Big[T_0(t)+\mathds{1}\,\tr\,T_0(t)\Big],
\end{align}
and
\beq
\label{eq:vector}
 \vec V(t)=\vec V^{\rm mat}(t)+\sqrt{2}G_F\Big[\vec T(t)+\mathds{1}\,\tr\,\vec T(t)\Big],
\eeq
where [see \Eqn{e:gmfbis}]
\beq
 \label{eq:rototo}
 T_0(t)=\int_{\vec p}\ell(t,\vec p)
\eeq 
and
\beq
\label{eq:totor}
 \vec T(t)=\int_{\vec p}\Big\{\hat p\,\ell(t,{\vec p})+\hat\epsilon_p\kappa(t,\vec p\,)+\hat\epsilon_p^*\kappa^\dagger(t,\vec p\,)\Big\}.
\eeq

The scalar and vector matter contributions in the active neutrino sector read, in the flavor basis,
\begin{align}
\label{eq:hmat}
h^{\rm mat}_{\alpha\beta}(t)&=\sqrt{2}G_F\delta_{\alpha\beta}\left[N_e(t)\delta_{\alpha e}-\frac{1}{2}N_n(t)\right],\\
\label{eq:Vmat}
\vec V^{\rm mat}_{\alpha\beta}(t)&=\sqrt{2}G_F\delta_{\alpha\beta}\left[\vec J_e(t)\delta_{\alpha e}-\frac{1}{2}\vec J_n(t)\right],
\end{align}
with the particle number and velocity densities ($\vec v_f=\vec p/E_p^f$)
\beq
 N_f(t)=2\int_{\vec p}\rho_f(t,\vec p\,)\quad {\rm and}\quad\vec J_f(t)=2\int_{\vec p}\vec v_f\rho_f(t,\vec p\,).
\eeq

Our final expression for the mean-field Hamiltonian in its $2N_f\times 2N_f$ matrix form thus reads
  \begin{equation}
  \label{eq:central}
{\cal H}(t, \vec q\,)= \left(
\begin{array}{cc}   
S(t,q)- \hat q\cdot \vec V(t) & -\hat\epsilon_q^*\cdot\vec V(t) \\
 -\hat\epsilon_q\cdot\vec V(t) &\bar S(t,q)+ \hat q\cdot \vec V(t)
\end{array}
\right).
\end{equation}
This is one of the main results of the present paper. We stress that nontrivial, off-diagonal particle-antiparticle mixing occur whenever the matrix $\vec V(t)\neq\vec 0$. Clearly, this requires anisotropic matter and/or neutrino backgrounds, as we mentioned earlier.\footnote{An obvious consequence of isotropy is that the correlators $\rho(t,\vec q\,)$, $\bar\rho(t,\vec q\,)$ and $\kappa(t,\vec q\,)$ only depend on $q=|\vec q|$. Isotropy further implies that $\kappa(t,q)=-\kappa^\dagger(t,q)$. For instance, this is needed to recover $\langle\bar\psi_i\vec \gamma\,(1-\gamma^5)\psi_j\rangle=\vec 0$.} We note finally that both the $\tr\,T_0(t)$ contribution to $S(t,q)$ and $\bar S(t,q)$ [Eqs. \eqn{eq:scalar} and \eqn{eq:scalarbar}] and the $N_n(t)$ contribution to $h^{\rm mat}(t)$ [\Eqn{eq:hmat}] give a term proportional to $\mathds{1}_{2N_f\times 2N_f}$ in the full mean-field Hamiltonian \eqn{eq:central} and can thus be discarded in the evolution equation \eqn{e:matrixform2}. This is, however, not the case for the $\tr\,\vec T(t)$ and $\vec J_n(t)$ contributions to the anisotropic term $\vec V(t)$ [\Eqn{eq:vector}] which give a nonzero off-diagonal neutrino-antineutrino coupling. It is easy to check explicitly that the corresponding Hamiltonian for Majorana neutrinos in the ultrarelativistic limit coincide with \Eqn{eq:central}, ${\cal H}_M(t,\vec q\,)={\cal H}(t,\vec q\,)$, as mentioned above.

We end this section by comparing \Eqn{eq:central} to the standard evolution Hamiltonian employed in studies of neutrino propagation in a supernova envelope. Assuming outward propagation of both particles and antiparticles from the neutrino sphere, such as in the bulb model \cite{Duan:2006an}, it is justified\footnote{It follows from the definitions \eqn{e:homogk} and \eqn{eq:nondiag1} that both the correlator $\kappa$ and the particle-antiparticle mixing $\Gamma^{\nu\bar\nu}$ require particles and antiparticles with (nearly) opposite momenta. For outward free-streaming, the spinors $u(\vec q\,)$ and $v(\vec q\,)$ in \Eqn{eq:nondiag1} should be understood to be nonzero only for outward momenta.} to set $\kappa\to0$ and $\Gamma^{\nu\bar\nu}\to0$ in Eqs. \eqn{eq:llmml}-\eqn{eq:vector}. Assuming an isotropic matter background such that $\vec J_f(t)=\vec 0$, one recovers the usual Hamiltonian\footnote{With our conventions the antineutrino number density matrix is $\bar\rho(t,-\vec q\,)$ and the corresponding Hamiltonian is thus obtained as $\Gamma^{\bar\nu\bar\nu}(t,-\vec q\,)|_{\kappa\to0}=-h^0(q)+h^{\rm mat}(t)+h^{\nu\nu}(t,\hat q)$, in accordance with known results \cite{Sigl:1992fn,Duan:2006an}.} 
\beq\label{e:hmf}
 \Gamma^{\nu\nu}(t,\vec q\,)\Big|_{\kappa\to0}=h^0(q)+h^{\rm mat}(t)+h^{\nu\nu}(t,\hat q),
\eeq
with\footnote{For $\Gamma^{\nu\bar\nu}\to0$, the particle and antiparticle sectors are decoupled and one can discard both the $\tr\,T_0(t)$ and $\tr\,\vec T(t)$ terms in the particle and antiparticle Hamiltonians $\Gamma^{\nu\nu}(t,\vec q\,)$ and $\Gamma^{\bar\nu\bar\nu}(t,\vec q\,)$.} 
\beq
\label{eq:nunustandard}
 h^{\nu\nu}(t,\hat q)=\sqrt{2}G_F\!\!\int_{\vec p}\left(1-\hat q\cdot\hat p\right)\ell(t,{\vec p}\,).
\eeq

We emphasize again that when the neutrino opacity cannot be neglected, simply assuming that $\kappa\to0$ is not a consistent approximation. Indeed, in such cases the particle-antiparticle mixing term in \Eqn{eq:central} is nonzero:
\beq\label{e:nterm}
\Gamma^{\nu\bar\nu}(t,\vec q\,)\Big|_{\kappa\to0}=-\sqrt{2}G_F\!\!\int_{\vec p}\left(\hat\epsilon_q^*\cdot\hat p\right)\Big[\ell(t,{\vec p}\,)+\mathds{1}\,\tr\,\ell(t,\vec p\,)\Big],
\eeq
from which it follows that nonzero pair correlations will be generated, as discussed previously.

Pair correlations where first discussed in Refs.~\cite{Volpe:2013uxl,Vaananen:2013qja} in the context of in-medium neutrino propagation. However, the class of mean-field dynamics studied there was not completely general in that it implemented particle and antiparticle number conservation separately and some terms included in the present general, fully relativistic treatment were missing. The evolution Hamiltonians in the particle and antiparticle sectors obtained in Refs.~\cite{Volpe:2013uxl,Vaananen:2013qja} were identical to Eqs.~\eqn{e:hmf} and \eqn{eq:nunustandard}, whereas the particle-antiparticle mixing term, referred to as the pairing (or abnormal) mean-field $\Delta$ in those articles, was given by\footnote{In particular, contrary to what happens in the general case discussed here, the limit $\kappa\to0$ is a consistent solution of the evolution equations derived in Refs.~\cite{Volpe:2013uxl,Vaananen:2013qja}.}
\beq
 \Delta(t,\vec q\,)=-\sqrt{2}G_F\!\!\int_{\vec p}\left(\hat\epsilon_q^*\cdot\hat\epsilon_p\right)\Big[\kappa(t,{\vec p}\,)+\mathds{1}\,\tr\,\kappa(t,{\vec p}\,)\Big].
\eeq
This corresponds to keeping only the third term in the general mean-field expression \eqn{eq:vector}.

\section{Helicity coherence}
\label{sec:hel}

We now turn to the study of neutrino-antineutrino mixing arising from nontrivial helicity coherence. These are encoded in the nondiagonal helicity components of the density correlators, e.g., $\rho_{-+}(t,\vec q\,)$ or $\bar \rho_{+-}(t,\vec q\,)$. Strictly speaking, these are related to neutrino-antineutrino mixing only in the case of Majorana neutrinos. For Dirac fields, such correlators involve a sterile component. Diagonal and nondiagonal helicity components are coupled through the nonvanishing neutrino masses and we thus consider the first nonrelativistic corrections to the evolution equations discussed in the previous section. For simplicity, we focus on helicity coherence throughout this section and set the pair correlations $\kappa\to0$ and the corresponding mixing terms in the evolution Hamiltonian $\Gamma^{\nu\bar\nu}\to0$. Nonrelativistic corrections in the presence of pair correlations are discussed in Appendix \ref{appsec:total}. As before, we first discuss the case of Dirac neutrinos and give the necessary modifications for the Majorana case.

\subsection{Mass corrections to the evolution equations}

In a general, spatially homogeneous system, the evolution equations for the equal-time two-point correlators keep the general structure \eqn{e:rhoevhom}--\eqn{e:kevhom} where all quantities are to be understood as $2N_f\times 2N_f$ matrices in mass/flavor and helicity space. For instance, keeping the mass/flavor indices implicit, the generalized particle  and antiparticle densities have the following helicity structure:\footnote{The $2N_f\times 2N_f$ matrix structure discussed here is to be distinguished from that of the previous section.}
\beq
\label{eq:helicity1}
 \rho(t,\vec q\,)\to
 \left(\begin{tabular}{cc}
 $\rho_{--}(t,\vec q\,)$&$\rho_{-+}(t,\vec q\,)$\\
 $\rho_{+-}(t,\vec q\,)$&$\rho_{++}(t,\vec q\,)$
\end{tabular}\right)\equiv
 \left(\begin{tabular}{cc}
 $\rho(t,\vec q\,)$&$\zeta(t,\vec q\,)$\\
 $\zeta^\dagger(t,\vec q\,)$&$\tilde\rho(t,\vec q\,)$
\end{tabular} \right)
\eeq
and 
\beq
\label{eq:helicity2}
 \bar\rho(t,\vec q\,)\to
   \left(\begin{tabular}{cc}
 $\bar\rho_{--}(t,\vec q\,)$&$\bar\rho_{-+}(t,\vec q\,)$\\
 $\bar\rho_{+-}(t,\vec q\,)$&$\bar\rho_{++}(t,\vec q\,)$
\end{tabular}  \right)\equiv
  \left(\begin{tabular}{cc}
 $\tilde{\bar\rho}(t,\vec q\,)$&$\bar\zeta^\dagger(t,\vec q\,)$\\
 $\bar\zeta(t,\vec q\,)$&$\bar\rho(t,\vec q\,)$
\end{tabular}  \right)\!,
\eeq
where the last equalities define our notation for the helicity correlators. For completeness we recall the definitions 
\begin{align}
 \rho_{ij,hh'}(t,\vec q\,)&=\langle a^\dagger_j(t,\vec q,h')a_i(t,\vec q,h)\rangle\\
 \bar\rho_{ij,hh'}(t,-\vec q\,)&=\langle b^\dagger_i(t,\vec q,h)b_j(t,\vec q,h')\rangle.
\end{align}
For instance, the helicity coherence term in the particle sector is $\zeta_{ij}(t,\vec q\,)=\langle a^\dagger_j(t,\vec q,+)a_i(t,\vec q,-)\rangle$, etc. Note that we use the same notation as in the previous section for the correlators $\rho_{ij}(t,\vec q\,)=\langle a^\dagger_j(t,\vec q,-)a_i(t,\vec q,-)\rangle$ and $\bar\rho_{ij}(t,-\vec q\,)=\langle b^\dagger_i(t,\vec q,+)b_j(t,\vec q,+)\rangle$. 

Similarly, the helicity components of the mean-field Hamiltonians in the particle and antiparticle sectors, i.e.,
\begin{align}
 \Gamma^{\nu\nu}_{ij,hh'}(t,\vec q\,)&=\bar u_i(\vec q,h)\tilde \Gamma_{ij}(t) u_j(\vec q,h')\\
 \Gamma^{\bar\nu\bar\nu}_{ij,hh'}(t,\vec q\,)&=\bar v_i(-\vec q,h)\tilde \Gamma_{ij}(t) v_j(-\vec q,h'),
\end{align}
can be gathered in a $2N_f\times 2N_f$ matrix form as
\beq
\label{eq:Hache}
 \Gamma^{\nu\nu}(t,\vec q\,)\!\to\!
   \left(\!\begin{tabular}{cc}
 $\Gamma^{\nu\nu}_{--}(t,\vec q\,)$&$\Gamma^{\nu\nu}_{-+}(t,\vec q\,)$\\
 $\Gamma^{\nu\nu}_{+-}(t,\vec q\,)$&$\Gamma^{\nu\nu}_{++}(t,\vec q\,)$
\end{tabular}\!  \right)\!\equiv\!
   \left(\!\!\begin{tabular}{cc}
 $H(t,\vec q\,)$&$\Phi(t,\vec q\,)$\\
 $\Phi^\dagger(t,\vec q\,)$&$\tilde H(t,\vec q\,)$
\end{tabular}\! \! \right)
\eeq
and
\beq
\label{eq:HacheBarre}
 \Gamma^{\bar\nu\bar\nu}(t,\vec q\,)\!\to\!
  \left(\!\begin{tabular}{cc}
 $\Gamma^{\bar\nu\bar\nu}_{--}(t,\vec q\,)$&$\Gamma^{\bar\nu\bar\nu}_{-+}(t,\vec q\,)$\\
 $\Gamma^{\bar\nu\bar\nu}_{+-}(t,\vec q\,)$&$\Gamma^{\bar\nu\bar\nu}_{++}(t,\vec q\,)$
\end{tabular}\! \right)\!\equiv\!
   \left(\!\begin{tabular}{cc}
 $\tilde{\bar H}(t,\vec q\,)$&$\bar\Phi^\dagger(t,\vec q\,)$\\
 $\bar\Phi(t,\vec q\,)$&$\bar H(t,\vec q\,)$
\end{tabular}\! \! \right)\!\!,
\eeq
where, again, the second equalities define our notations. 
Discarding pair correlations, the evolution equations read
\begin{align}
\label{eq:evolhelrho}
 i\dot\rho(t,\vec q\,)&=\left[\Gamma^{\nu\nu}(t,\vec q\,),\rho(t,\vec q\,)\right]\\
\label{eq:evolhelrhobar}
 i\dot{\bar\rho}(t,\vec q\,)&=\left[\Gamma^{\bar\nu\bar\nu}(t,\vec q\,),\bar\rho(t,\vec q\,)\right].
\end{align}

\subsection{Mass corrections to the mean-field Hamiltonian}

Now we compute the  ${\cal O}(m/q)$ corrections to the interaction part of the mean-field Hamiltonian. The matter kernel \eqn{eq:matterham} is the same as in the previous section, whereas the contribution \eqn{eq:mmm} from the neutrino self-interaction receives corrections from the off-diagonal helicity components of the various density matrices. A straightforward calculation of spinor products (see Appendix~\ref{appsec:spinors})  yields
\beq
 \label{e:gmf2}
 \tilde\Gamma^{\rm self}(t)={G_F \over{\sqrt{2}}}\gamma^{\mu}(1-\gamma_5) \Big[T_\mu(t)+\mathds{1}\,\tr\,T_\mu(t)\Big],
\eeq
with
\begin{align}
\label{e:gmf2bis}
 T_\mu (t)&=  \int_{\vec p} \Big\{ n_\mu(\hat p)\ell(t,\vec p\,)\nn
  &- e^{-i\phi_p}\epsilon_\mu(\hat p)\Omega(t,\vec p\,)\frac{m}{2p}-   \frac{m}{2p}e^{i\phi_p}\epsilon_\mu^*(\hat p)\Omega^\dagger(t,\vec p\,) \Big\},
\end{align}
where $\phi_p$ is the polar angle of the vector $\hat p$ in spherical coordinates. Here, $m$ denotes the neutrino mass matrix and we introduced 
the combination 
\beq
\label{eq:omegadef}
 \Omega(t,\vec p\,)=\zeta(t,\vec p\,)+\bar\zeta(t,-\vec p\,).
\eeq

Finally, we obtain, for the components of the mean-field Hamiltonians \eqn{eq:Hache} and \eqn{eq:HacheBarre} up to contributions ${\cal O}(m^2/q^2)$ for the interaction terms,\footnote{Here, we only keep the corrections ${\cal O}( m/q)$ in the interaction terms. The free contribution $h^0(q)$ can be expanded independently $h^0(q)\approx q\mathds{1}+m^2/2q$. Note that the leading-order kinetic term does not contribute to the evolution equations \eqn{eq:evolhelrho} and \eqn{eq:evolhelrhobar} and can thus be discarded. As noticed previously, this is not possible when pair correlations are present.}
\begin{align}
\label{eq:hamcompfirst}
 H(t,\vec q\,)&=S(t,q)-\hat q\cdot\vec V(t)-\hat q\cdot \vec V_m(t),\\
 \Phi(t,\vec q\,)&=e^{i\phi_q}\hat\epsilon_q^*\cdot\vec V(t){m\over 2q},\\
 \tilde H(t,\vec q\,)&=h^0(q),
\end{align}
and
\begin{align}
\label{eq:hamcompfirstbar}
 \bar H(t,\vec q\,)&=\bar S(t,q)+\hat q\cdot\vec V(t)+\hat q\cdot \vec V_m(t),\\
 \bar\Phi(t,\vec q\,)&=e^{i\phi_q}\hat\epsilon_q\cdot\vec V(t){m\over 2q},\\
\label{eq:hamcomplast}
 \tilde{\bar H}(t,\vec q\,)&=-h^0(q),
\end{align}
where $S(t,q)$, $\bar S(t,q)$ and $\vec V(t)$ have been defined in Eqs. \eqn{eq:scalar}, \eqn{eq:scalarbar}, and \eqn{eq:vector}, respectively\footnote{It is understood here that  Eqs.~\eqn{eq:scalar}, \eqn{eq:scalarbar}, and \eqn{eq:vector} are evaluated at $\kappa=0$; see Appendix \ref{appsec:total} for the generalization to $\kappa\neq0$.} and where the mass correction to the vector component of the mean-field reads
\beq
\label{eq:masscorrecvec}
 \vec V_m(t)=\sqrt{2}G_F\Big[\vec T_m(t)+\mathds{1}\,\tr\,\vec T_m(t)\Big],
\eeq
with
\beq
\label{eq:masscorrecvecbis}
 \vec T_m(t)=-\int_{\vec p} \Big\{ e^{-i\phi_p}\hat\epsilon_p\,\Omega(t,\vec p\,)\frac{m}{2p}  +  {\rm H.c.}\Big\}.
\eeq

As expected, the off-diagonal terms in the particle and antiparticle Hamiltonians, $\Phi$ and $\bar\Phi$, are proportional to the neutrino masses and the different helicity sectors are decoupled in the ultrarelativistic limit.

\subsection{Majorana neutrinos}

As before, the treatment of Majorana neutrinos closely follows that of Dirac neutrinos. One introduces a similar matrix in helicity as in Eqs. \eqn{eq:helicity1} and \eqn{eq:Hache}, where $\tilde \rho_M(t,\vec q\,)=\bar \rho^T_M(t,-\vec q\,)$ and $\tilde H_M(t,\vec q\,)=-\bar H^T_M(t,-\vec q\,)$; see Eqs. \eqn{eq:Mrel} and \eqn{eq:MM}. We thus write
\beq
 \rho_M(t,\vec q\,)\to
 \left(\begin{tabular}{cc}
 $\rho_M(t,\vec q\,)$&$\zeta_M(t,\vec q\,)$\\
 $\zeta^\dagger_M(t,\vec q\,)$&$\bar \rho_M^T(t,-\vec q\,)$
\end{tabular} \right)
\eeq
and 
\beq
\label{eq:HacheMajo}
 \Gamma_M^{\nu\nu}(t,\vec q\,)\to
   \left(\begin{tabular}{cc}
 $H_M(t,\vec q\,)$&$\Phi_M(t,\vec q\,)$\\
 $\Phi^\dagger_M(t,\vec q\,)$&$-\bar H^T_M(t,-\vec q\,)$
\end{tabular} \right).
\eeq
Using the results of the previous section together with Eqs.~\eqn{eqn:MajoHam1}--\eqn{eqn:MajoHam4}, we obtain
\begin{align}
\label{eq:hamcompfirstMajo}
 H_M(t,\vec q\,)&=S(t,q)-\hat q\cdot\vec V(t)-\hat q\cdot \vec V_m(t),\\
 \bar H_M(t,\vec q\,)&=\bar S(t,q)+\hat q\cdot\vec V(t)+\hat q\cdot \vec V_m(t),\\
 \Phi_M(t,\vec q\,)&=e^{i\phi_q}\hat \epsilon^*_q\cdot\left[\vec V(t)\frac{m}{2q}+\frac{m}{2q}\,{\vec V}^T\!(t)\right].
\end{align}

We have explicitly checked that the evolution equations derived here reproduce\footnote{In terms of the four-vectors $x_{1,2}^\mu$ introduced in Refs.~\cite{Vlasenko:2013fja,Cirigliano:2014aoa}, we have $e^{-i\phi_p}\epsilon^\mu(\hat p)=x_1^\mu-ix_2^\mu$; see also Appendix \ref{appsec:total}. In the Dirac case, we find a different sign for the off-diagonal term $\bar\Phi$ of the antiparticle Hamiltonian as compared to Ref.~\cite{Cirigliano:2014aoa}.} those obtained in Refs. \cite{Vlasenko:2013fja,Cirigliano:2014aoa}, except for the mass correction \eqn{eq:masscorrecvec} to the diagonal helicity components of the various evolution Hamiltonians, which has only been given in an implicit form in those references.\footnote{The authors of Refs.~\cite{Vlasenko:2013fja,Cirigliano:2014aoa} have privately confirmed that their expressions agree with \Eqn{eq:masscorrecvec}.} 

\section{Conclusions}
\label{sec:concl}

Investigating neutrino flavor evolution in dense astrophysical environments has numerous fascinating facets. In the present work, we have derived the most general mean-field equations that are applicable in the case of inhomogeneous and anisotropic neutrino and matter backgrounds, for either massive Dirac or Majorana neutrinos. These encompass all mean-field evolution equations derived previously. 

We have emphasized the presence of pairing correlations between neutrinos and antineutrinos and, for the Majorana case, also between pairs of neutrinos or antineutrinos. Such pair correlations---which correspond to the fermionic analog of squeezed bosonic states---were first discussed in the context of neutrino physics in Refs.~\cite{Volpe:2013uxl,Vaananen:2013qja}. Our results generalize the evolution equation presented in those references to the case of a completely general, fully relativistic mean-field Hamiltonian. 

We have also considered possible nontrivial helicity correlations due to the nonvanishing neutrino masses. These can be of either the normal or the pairing type. The former case corresponds to helicity coherence terms, which were discussed in Refs.~\cite{Vlasenko:2013fja,Cirigliano:2014aoa,Vlasenko:2014bva}, whereas the latter were first pointed out in Ref.~\cite{Volpe:2013uxl}. We have treated such correlations by systematically including the first nonrelativistic corrections to the general mean-field evolution equations.  When pair correlations are neglected, our results reproduce the evolution equations of Refs.~\cite{Vlasenko:2013fja,Cirigliano:2014aoa}. The latter were implemented in a schematic one-flavor calculation in Ref.~\cite{Vlasenko:2014bva}, which showed that helicity correlations can be of importance under specific conditions.

Finally, we have considered the full nontrivial helicity structure of the general mean-field equations, including pair correlations, at leading order in the nonvanishing neutrino masses. This couples the effects of helicity coherence and pair correlations. Both effects may be of importance in dense anisotropic environments with nonzero neutrino opacity, e.g., the interior of the neutrinosphere in core-collapse supernov{ae}. Future numerical investigations either in schematic models or in realistic simulations will reveal whether the various types of neutrino-antineutrino mixings can influence neutrino flavor evolution and possibly impact the nucleosynthetic outcomes in stellar environments, or the shock revival in supernov{ae}. 

Although mean-field calculations are an important step in assessing the possible role of unusual correlations, a more complete treatment of the dense astrophysical environments of interest here would require one to include collision terms. Formal kinetic equations including helicity coherence for neutrinos in anisotropic environments were discussed in Refs.~\cite{Vlasenko:2013fja,Cirigliano:2014aoa}, although the definite form of the collision term still needs to be derived. On the other hand, a general formalism to consistently derive kinetic equations that take proper account of pair correlations was developed in Refs.~\cite{Herranen:2010mh,Fidler:2011yq} for isotropic systems. This has been applied to a model system of fermionic and scalar fields with a Yukawa interaction in the context of lepto-/baryogenesis. It would be of interest to apply these methods to neutrinos in situations of astrophysical interest.

\appendix

\section{Complete mean-field Hamiltonian at~${\cal O}(m/q)$}
\label{appsec:total}

Here we present the general mean-field equations for spatially homogeneous systems with both pair correlations and helicity coherence terms, including nonrelativistic  corrections to the interaction Hamiltonian at first order in the neutrino masses. As explained in Sec. \ref{sec:hel}, the equations of motion keep the same form as Eqs.~\eqn{e:rhoevhom}--\eqn{e:kevhom}, where each term is a $2N_f\times 2N_f$ matrix in mass/flavor and helicity space. Equivalently, these can be written as \Eqn{e:matrixform2} where both the generalized density matrix \eqn{e:genR2} and Hamiltonian \eqn{e:genH2} are $4N_f\times4N_f$ matrices. The diagonal blocks have been discussed in Sec. \ref{sec:hel}; see Eqs. \eqn{eq:helicity1}, \eqn{eq:helicity2}, \eqn{eq:Hache}, and \eqn{eq:HacheBarre}. Similarly, we write the helicity structure of the off-diagonal blocks as
\beq
 \kappa(t,\vec q\,)\to
 \left(\begin{tabular}{cc}
 $\kappa_{--}(t,\vec q\,)$&$\kappa_{-+}(t,\vec q\,)$\\
 $\kappa_{+-}(t,\vec q\,)$&$\kappa_{++}(t,\vec q\,)$
\end{tabular}\right)\equiv
\left(\begin{tabular}{cc}
 $\varkappa(t,\vec q\,)$&$\kappa(t,\vec q\,)$\\
 $\kappa^\dagger(t,\vec q\,)$&$\tilde{\!\varkappa}(t,\vec q\,)$
\end{tabular}\right)
\eeq
for pairing correlations (the second equality defines our notation), and
\beq
\label{appeq:gammanunubar}
 \Gamma^{\nu\bar\nu}(t,\vec q\,)\to
   \left(\begin{tabular}{cc}
 $\Gamma^{\nu\bar\nu}_{--}(t,\vec q\,)$&$\Gamma^{\nu\bar\nu}_{-+}(t,\vec q\,)$\\
 $\Gamma^{\nu\bar\nu}_{+-}(t,\vec q\,)$&$\Gamma^{\nu\bar\nu}_{++}(t,\vec q\,)$
\end{tabular} \right)
\eeq
for the particle-antiparticle mixing terms in the Hamiltonian.

To discuss the explicit form of the mean-field Hamiltonian at ${\cal O}(m/q)$, we introduce the following notation for the interaction part:
\beq
 \tilde \Gamma^{\rm int}(t)=\frac{1}{2}\gamma^\mu(1-\gamma_5)\Sigma_\mu(t),
\eeq
where the $N_f\times N_f$ matrices\footnote{These are the right-handed vector components of the neutrino self-energy in the mean-field approximation \cite{Vlasenko:2013fja,Cirigliano:2014aoa}.} $\Sigma_\mu(t)$ can be decomposed into a matter part $\Sigma_\mu^{\rm mat}(t)$ and a contribution from neutrino self-interactions of the general form  
\begin{align}
 \label{eq:sigmamu}
 \Sigma_\mu(t)= \Sigma^{\rm mat}_\mu(t)+\sqrt{2}G_F\Big[T_\mu(t)+\mathds{1}\,\tr\,T_\mu(t)\Big],
\end{align}
with
\begin{align}
 T_\mu(t)=\int_{\vec p}\Big\{&n_\mu(\hat p) \left[{\cal R}(t,\vec p\,)-\bar {\cal R}(t,-\vec p\,)\right]\nn
 +&\epsilon_\mu(\hat p) {\cal K}(t,\vec p\,)+\epsilon_\mu^*(\hat p){\cal K}^\dagger(t,\vec p\,)\Big\},
\end{align}
where ${\cal R}$, $\bar {\cal R}$, and ${\cal K}$ are combinations of the various two-point correlators. The matter contribution has been considered in Sec. \ref{sec:pair} and reads, in the flavor basis,
\beq
 \Sigma_{\mu,\alpha\beta}^{\rm mat}(t)=\sqrt{2}G_F\delta_{\alpha\beta}\Big[J_\mu^e(t)\delta_{\alpha e}+\sum_{f=e,p,n}c_V^f J_\mu^f(t)\Big],
\eeq
where $c_V^e=-c_V^p=2\sin^2\theta_W-1/2$ and $c_V^n=-1/2$, where $\theta_W$ is the Weinberg angle.

Using the spinor products recalled in Appendix \ref{appsec:spinors}, a straightforward calculation of the correlator in \Eqn{eq:mmm} yields, at first nontrivial order in the neutrino mass matrix $m$,
\begin{align}
 {\cal R}(t,\vec p\,)&=\rho(t,\vec p\,)-e^{-i\phi_p}\varkappa(t,\vec p\,)\frac{m}{2p}-\frac{m}{2p}e^{i\phi_p}\varkappa^\dagger(t,\vec p\,),\\
 \bar {\cal R}(t,\vec p\,)&=\bar \rho(t,\vec p\,)+e^{-i\phi_p}\,\tilde{\!\varkappa}^\dagger(t,\vec p\,)\frac{m}{2p}+\frac{m}{2p}e^{i\phi_p}\,\tilde{\!\varkappa}(t,\vec p\,),\\
 {\cal K}(t,\vec p\,)&=\kappa(t,\vec p\,)-e^{-i\phi_p}\Omega(t,\vec p\,)\frac{m}{2p},
\end{align}
where $\Omega(t,\vec p\,)$ has been defined in \Eqn{eq:omegadef}.

\subsection{Dirac case}

Using the spinor products of Appendix \ref{appsec:spinors} again and systematically retaining ${\cal O}(m/q)$ terms from the interaction parts, we finally obtain the components of the mean-field Hamiltonian as
\begin{align}
 \Gamma^{\nu\nu}_{--}(t,\vec q\,)&=h^0(q)+n_\mu(\hat q)\Sigma^\mu(t),\\
 \Gamma^{\nu\nu}_{-+}(t,\vec q\,)&=-e^{i\phi_q}\epsilon_\mu^*(\hat q)\Sigma_0^\mu(t)\frac{m}{2q},\\
 \Gamma^{\nu\nu}_{+-}(t,\vec q\,)&=-e^{-i\phi_q}\frac{m}{2q}\epsilon_\mu(\hat q)\Sigma_0^\mu(t),\quad\\
\Gamma^{\nu\nu}_{++}(t,\vec q\,)&=h^0(q)
\end{align}
for the particle-particle sector,
\begin{align}
 \Gamma^{\bar\nu\bar\nu}_{--}(t,\vec q\,)&=-h^0(q),\\
 \Gamma^{\bar\nu\bar\nu}_{-+}(t,\vec q\,)&=-e^{-i\phi_q}\frac{m}{2q}\epsilon_\mu^*(\hat q)\Sigma_0^\mu(t),\\
 \Gamma^{\bar\nu\bar\nu}_{+-}(t,\vec q\,)&=-e^{i\phi_q}\epsilon_\mu(\hat q)\Sigma_0^\mu(t)\frac{m}{2q},\\
 \Gamma^{\bar\nu\bar\nu}_{++}(t,\vec q\,)&=-h^0(q)+n_\mu(-\hat q)\Sigma^\mu(t)
\end{align}
for the antiparticle-antiparticle sector, and
\begin{align}
 \Gamma^{\nu\bar\nu}_{--}(t,\vec q\,)&=-e^{i\phi_q}n_\mu(\hat q)\Sigma_0^\mu(t)\frac{m}{2q},\\
 \Gamma^{\nu\bar\nu}_{-+}(t,\vec q\,)&=\epsilon^*_\mu(\hat q)\Sigma^\mu(t),\\
 \Gamma^{\nu\bar\nu}_{+-}(t,\vec q\,)&=0,\\
 \Gamma^{\nu\bar\nu}_{++}(t,\vec q\,)&=-e^{-i\phi_q}\frac{m}{2q}n_\mu(-\hat q)\Sigma_0^\mu(t)
\end{align}
for the particle-antiparticle mixing terms. The components of $\Gamma^{\bar\nu\nu}$ are obtained from \Eqn{eq:ksfdasdf}. Here $\Sigma_0^\mu(t)$ is obtained using \Eqn{eq:sigmamu} evaluated at vanishing mass. This provides a complete set of evolution equations up to (and including) ${\cal O}(G_F m/q)$ contributions, thus generalizing the equations presented in Secs.~\ref{sec:pair} and \ref{sec:hel}.

\subsection{Majorana case}

We obtain the independent component of the mean-field Hamiltonian for Majorana neutrinos from the results of the previous subsection by means of E	qs.~\eqn{eqn:MajoHam1} and \eqn{eqn:MajoHam3}. With our conventions \eqn{e:g1hom}--\eqn{e:g4hom} for the momentum assignments in the case of spatially homogeneous systems, these become
\begin{align}
 \Gamma^{\nu\nu}_M(t,\vec q\,)&=\Gamma^{\nu\nu}(t,\vec q\,)-\left[\Gamma^{\bar\nu\bar\nu}(t,-\vec q\,)\right]^T,\\
 \Gamma^{\nu\bar\nu}_M(t,\vec q\,)&=\Gamma^{\nu\bar\nu}(t,\vec q\,)-\left[\Gamma^{\nu\bar\nu}(t,-\vec q\,)\right]^T,
\end{align}
where the transposition acts on both mass/flavor and helicity indices. Keeping in mind the different normalization of the free contribution to the Hamiltonian in the Majorana case, i.e., $h^0(q)\to h^0(q)/2$, we obtain 
\begin{align}
 \Gamma^{\nu\nu}_{M,--}(t,\vec q\,)&=h^0(q)+n^\mu(\hat q)\Sigma_\mu(t),\\
 \Gamma^{\nu\nu}_{M,-+}(t,\vec q\,)&=-e^{i\phi_q}\epsilon^{\mu*}(\hat q)\!\left[\Sigma_{0\mu}(t)\frac{m}{2q}+\frac{m}{2q}\Sigma_{0\mu}^T(t)\right],\\
 \Gamma^{\nu\nu}_{M,+-}(t,\vec q\,)&=-e^{-i\phi_q}\epsilon^\mu(\hat q)\!\left[\frac{m}{2q}\Sigma_{0\mu}(t)+\Sigma_{0\mu}^T(t)\frac{m}{2q}\right]\quad,\\
\Gamma^{\nu\nu}_{M,++}(t,\vec q\,)&=h^0(q)-n^\mu(\hat q)\Sigma^T_{\mu}(t),
\end{align}
and 
\begin{align}
 \Gamma^{\nu\bar\nu}_{M,--}(t,\vec q\,)&\!=\!-e^{i\phi_q}\!\!\left[n_\mu(\hat q)\Sigma_{0}^{\mu}(t)\frac{m}{2q}+\frac{m}{2q}n^\mu(-\hat q)\Sigma_{0\mu}^T(t)\right],\\
 \Gamma^{\nu\bar\nu}_{M,-+}(t,\vec q\,)&=\epsilon^*_\mu(\hat q)\Sigma^\mu(t),\\
 \Gamma^{\nu\bar\nu}_{M,+-}(t,\vec q\,)&=-\epsilon^\mu(\hat q)\Sigma^T_\mu(t),\\
 \Gamma^{\nu\bar\nu}_{M,++}(t,\vec q\,)&\!=\!-e^{-i\phi_q}\!\!\left[\frac{m}{2q}n_\mu({\rm -}\hat q)\Sigma_{0}^{\mu}(t)\!+\!n^\mu(\hat q)\Sigma^T_{0\mu}(t)\frac{m}{2q}\right]\!\!.
\end{align}

\section{Spinor products}
\label{appsec:spinors}

We use the following representation for Dirac bispinors in the chiral representation \cite{Giunti:2007ry}:
\begin{align}
 u_i(\vec p,h)&=\left(\begin{tabular}{c}
 $-{\cal N}_{p,-h}^i\,\chi^h(\hat p)$\\\vspace*{-.35cm}\\${\cal N}_{p,h}^{i}\,\chi^h(\hat p)$
 \end{tabular}\right),\\
 v_i(\vec p,h)&=-h\left(\begin{tabular}{c}
 ${\cal N}_{p,h}^{i}\,\chi^{-h}(\hat p)$\\\vspace*{-.35cm}\\${\cal N}_{p,-h}^{i}\,\chi^{-h}(\hat p)$
 \end{tabular}\right),
\end{align}
with the standard two-component helicity spinors 
\beq
 \chi^+(\hat p)=\left(\begin{tabular}{c}
 $\cos{\theta_p\over2}$\\$e^{i\phi_p}\sin{\theta_p\over2}$
 \end{tabular}\right)\,,\quad\chi^-(\hat p)=\left(\begin{tabular}{c}
 $-e^{-i\phi_p}\sin{\theta_p\over2}$\\$\cos{\theta_p\over2}$
 \end{tabular}\right),
\eeq
and where we have defined ($\varepsilon_{i,p}=\sqrt{p^2+m_i^2}$)
\beq
 {\cal N}_{p,h}^i=\sqrt{\frac{\varepsilon_{i,p}-hp}{2\varepsilon_{i,p}}}=\delta_{h-}+\frac{m_i}{2p}\delta_{h+}+{\cal O}(m_i^2/p^2).
\eeq
Introducing the notation
\beq
 \gamma_L^\mu=\gamma^\mu\frac{1-\gamma_5}{2},
\eeq
the relevant bispinor products read
\begin{align}
 \bar u_j(\vec p,h)\gamma^\mu_L\,u_i(\vec p,h)&={\cal N}_{p,h}^j\,{\cal N}_{p,h}^i\,n^\mu(-h\hat p)\approx n^\mu(\hat p)\delta_{h-},\\
 \bar v_j(-\vec p,-h)\gamma^\mu_L\,u_i(\vec p,h)&={\cal N}_{p,h}^{j}\,{\cal N}_{p,h}^{i}\,\epsilon^\mu(-h\hat p)\approx\epsilon^\mu(\hat p)\delta_{h-},
\end{align}
and
\begin{align}
 \bar u_j(\vec p,-h)&\gamma^\mu_L\,u_i(\vec p,h)=-{\cal N}_{p,-h}^{j}\,{\cal N}_{p,h}^{i}\,e^{ih\phi_p}\epsilon^\mu(-h\hat p)\nn
 &\approx-\frac{m_j}{2p}e^{-i\phi_p}\epsilon^\mu(\hat p)\delta_{h-}-\frac{m_i}{2p}e^{i\phi_p}\epsilon^{\mu*}(\hat p)\delta_{h+},\\
 \bar v_j(-\vec p,h)&\gamma^\mu_L\,u_i(\vec p,h)=-{\cal N}_{p,-h}^{j}\,{\cal N}_{p,h}^{i}\,e^{ih\phi_p}n^\mu(-h\hat p)\nn
 &\approx-\frac{m_j}{2p}e^{-i\phi_p}n^\mu(\hat p)\delta_{h-}-\frac{m_i}{2p}e^{i\phi_p}n^{\mu}(-\hat p)\delta_{h+},
\end{align}
where the approximate expressions are valid up to relative corrections of order $m^2/p^2$. The other bispinor products needed in the text can be obtained from the relations 
\beq
 \bar v_j(\vec p,h)\gamma^\mu_L\,v_i(\vec p,h')=hh' \,\bar u_j(\vec p,-h)\gamma^\mu_L\,u_i(\vec p,-h')
\eeq
and
\beq
 \bar u_j(\vec p,h)\gamma^\mu_L\,v_i(-\vec p,h')=\left[\bar v_i(-\vec p,h')\gamma^\mu_L\,u_j(\vec p,h)\right]^*.
\eeq

\end{document}